# Substrate surface engineering for tailoring properties of functional ceramic thin films

H. -U. Habermeier[1]

[1] Max-Planck-Institut für Festkörperforschung , Heisenbergstr.1 D 70569 Stuttgart

Using oxide substrates for functional ceramic thin film deposition beyond their usual application as chemical inert, lattice-matched support for the films represents a novel concept in ceramic thin film research. The substrates are applied as a functional element in order to controllably modify the atom arrangement and the growth mode of ceramic prototype materials such as cuprate superconductors and colossal magnetoresistance manganites. One example is the use of epitaxial strain to adjust the relative positions of cations and anions in the film and thus modify their physical properties. The other makes use of vicinal cut $SrTiO_3$ which enables the fabrication of regular nanoscale step and terrace structures. In $YBa_2Cu_3O_{7-x}$ thin films grown on vicinal cut $SrTiO_3$ single crystals a regular array of antiphase boundaries is generated causing an anisotropic enhancement of flux-line pinning. In the case of La-Ca-Mn-O thin films grown on vicinal cut substrates it could be demonstrated that magnetic in-plane anisotropy is achieved.

## 1 Introduction

The physical properties of perovskite-type functional ceramic thin films are known to sensitively depend on details of deviations from their ideal composition and/or crystal structure. Therefore, substrate-induced lattice strain, substrate surface morphology and growth-induced defects are some examples for extrinsic effects playing an important role in determining thin-film properties and consequently the application potential of functional ceramics [1-5]. The sensitivity of the physical properties on the structure arises from the origin of the functionality at the level of sub-unit cells. In the case of the doped rare earth manganites, e.g. the bonding distance and bonding angle of the Mn-O-Mn building block determines the charge transfer of an electron from Mn to Mn via the oxygen, thus the bandwidth, metallicity, charge-ordering and the appearance of ferromagnetism [6,7,8]. In high temperature superconductor [HTS] cuprates such as $La_{1-x}Sr_xCuO_4$ the distance of the apex oxygen from the $CuO_2$ planes affects the Cu4s-O2p hybridization and thus doping and $T_C$ [9,10].

The multi-component chemical composition of the ceramic materials combined with their complex crystal structure represents a much higher degree of sophistication for a thin-film technology compared to that of metals and semiconductors. Consequently, to understand the film growth process and modify it intentionally in order to open a path for defect control, is a tremendous challenge. In contrast to metals and semiconductors where the deposition temperature for epitaxial growth from the vapor phase is around 20% (metals) to 40% (semiconductors) of the melting temperature, $T_M$, for ceramics such as the HTS cuprates much higher values of around (0.7 – 0.8) $T_M$ are required. This reflects the chemical dissimilarity of the constituent cations and their quite different diffusion coefficient at the substrate surface at growth conditions. The general problem of the vapor deposition of oxides with complex chemical composition and large unit cells to fabricate single crystal type epitaxially grown thin films had not been addressed prior to 1986, the year of the discovery of the HTS cuprates. The role of the substrates in the efforts exploring the epitactic growth of ceramic thin films has been treated so far simply as that of a mechanical support combined with chemical stability and compatibility with the prerequisites given by the epitaxy relations. In order to pave a new way for all-oxide electronics and novel device concepts an advanced oxide epitaxy technology is required based on nanoscale substrate engineering as well as on atomic layer control of oxide films. Furthermore, the opportunities buried in tailoring the substrate surface morphology have to be explored in order to intentionally modify growth conditions and defect arrangements.

In this paper, a novel concept in ceramic thin-film research is introduced, the use of substrates as a functional element. The function of a substrate can originate either from well designed topological properties affecting the growth and defect structure of a ceramic thin film or from its electronic and/or magnetic properties. In a first step, we restrict ourselves to functions derived from the substrate surface and structural peculiarities of artificially tailored substrates. Two examples may serve for this case study. First, an intentional lattice mismatch is used to tailor the epitaxial strain of the film, second, vicinal cut surfaces with a nanoscale step-and-terrace structure are applied to generate macroscopically aligned antiphase boundaries

## 2. Some basic considerations for ceramic thin film growth

The key requirements of a mature technology for future device applications of functional ceramics are the ability to deposit compact homogeneous layers and multilayer structures with flat surfaces in the view of the short correlation lengths e.g. for superconducting or ferromagnetic thin films. The difficulties achieving structurally and morpholologically perfect ceramic thin films of a specific type are connected with the complex parameter space for deposition spanned by deposition rate, kinetic energy of the particles impinging the substrate surface, surface energies of substrate, film and interface, respectively, substrate flatness and termination. Additionally, the thermo-dynamic requirements for phase stability as given e.g. in the oxygen-pressure/ temperature phase diagram [11] for HTS thin film deposition has to be fulfilled. In general, the rate of growth and the morphology of a film depend on several factors. The most important one, governing the nucleation and growth mode, is the relative supersaturation, $\mu$, determining the chemical potential, $\sigma$, as driving force for epitaxy. They are related by

$$\sigma = k_B T_s \ln\mu = k_B T_s \ln(\Phi \Delta T / R T_s^2) \qquad (1)$$

where $k_B$ is the Boltzmann constant, $\Phi$ denotes the molar heat of solution, $\Delta T$ the undercooling, R and $T_s$ are the gas constant and the absolute temperature of the substrate. When the thermodynamic driving force for crystallization from the vapor phase is small, (small supersaturation) dislocation-controlled growth (spiral growth) is observed. The growth mechanism is spiral type in which nuclei are continuously added to the edge of a spirally expanding step. Surface steps are energetically favorable attachment sites. Growth of materials with highly anisotropic surface energies and thus growth velocities such as $YBa_2Cu_3O_{7-x}$ [growth velocity in *a, b*-direction much larger than in *c*-direction] has an extended region for spiral growth. Increase of the effective supersaturation favors two-dimensional nucleation [island formation] and a crossover from dislocation-dominated to nucleation-dominated growth is expected at a critical supersaturation $\mu^*$. Further increase of $\mu$ causes a transition to unstable growth, uncontrolled nucleation of growth centers on top of each others lead to a dendritic growth type [12]. These regimes are schematically shown in Fig. 1. In addition to supersaturation the deposition temperature, $T_s$, plays an important role in determining the growth kinetics and the surface morphology. Defining a normalized bonding energy, $E_b$, for the substrate surface atoms

$$E_b = 4\Phi_{ss}/2k_BT_s \qquad (2)$$

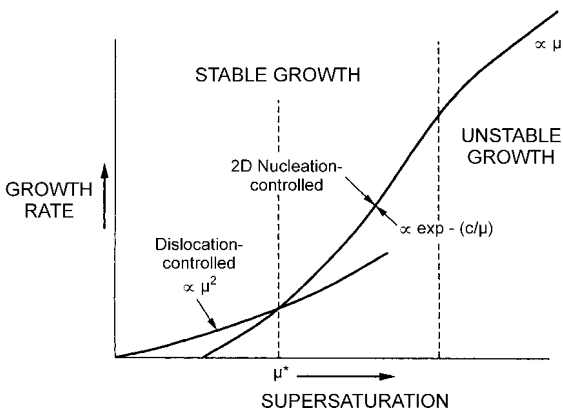

Fig. 1. Schematic dependence of the growth regimes on super-saturation (taken from Ref. 12 ex-planations are given in the text)

($\Phi_{ss}$ denotes the potential energy of a solid–solid nearest neighbor pairs of atoms in the substrate unit cell) it is obvious that increasing deposition temperature implies a smaller $E_b$ and the surface becomes rougher. Thus a higher density of kink sites on the surface is offered for the oncoming vapor particles, thus leading to a more rapid growth. For the spontaneous nucleation of a unit cell a critical volume of the deposited material is required. Surface diffusion that supplies a nucleus with the necessary material, however, is quite different for the cationic constituents of the material. For $YBa_2Cu_3O_{7-x}$ e.g. the surface diffusion coefficients at deposition temperature, $T_s = 800^0$ C, vary by 4 orders of magnitude from Y ($10^{-13}$ m$^2$/s) to Cu ($10^{-9}$ m$^2$/s), thus facilitating the formation of micro-precipitates and a phase separation if the stoichiometry is not perfect [13].

### 3. The "perfect" film on a "perfect" substrate

As briefly sketched in the previous section, supersaturation and substrate temperature determine whether the nucleation is dislocation controlled or island growth controlled. Substrates for the growth of ceramic thin films are more than just a mechanical support for the film. The ideal substrate has to fulfill the requirements of perfect lattice match to ensure epitaxy, match of the thermal expansion coefficients to avoid cracking, lack of structural phase transition between deposition temperature and operating temperature to pre-vent additional stress, chemical inertness with respect to the film forming species and, finally, lack of interdiffusion. The misfit between the substrate and the film at epitaxial growth temperature not only affects the selection of the epitaxy relations it also influences the surface nucleation and growth modes.

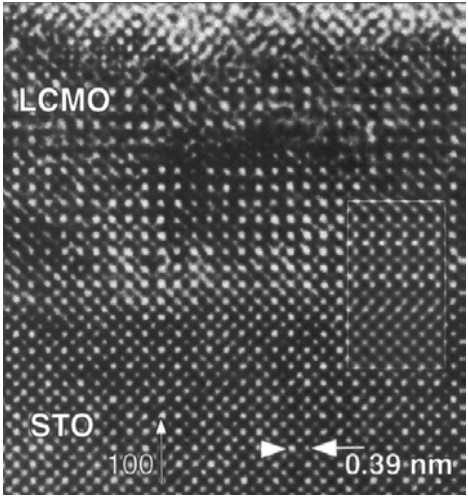

Fig. 2 High resolution cross-sectional TEM of an $SrTiO_3/La_{2/3}Ca_{1/3}MnO_3$ interface showing perfect epitaxy.

Misfit reduces the step-flow regime and enhances two-dimensional nucleation. Additionally, the stored elastic energy in the pseudomorphically grown layer adjacent to the substrate acts as a driving force for cluster generation. Misfit acts in the same direction as supersaturation. Strain caused by lattice mismatch and cooling due to thermal expansion coefficient differences can be accommodated by misfit dislocation formation, twinning, and redistribution of oxygen in the lattice. If these mechanisms cannot relieve the strain, cracking may occur. The difficulties in achieving the "perfect" ceramic thin film are connected with problems of the phase stability and oxygen supply superimposed to the peculiarities of the kinetically controlled thin film growth process and the substrate related effects. In Fig. 2 a TEM cross section micrograph of a $La_{2/3}Ca_{1/3}MnO_3$ thin film is given showing the perfect epitaxy on the $SrTiO_3$ substrate [14].

## 4. Tailoring epitaxial strain in HTS thin films

It is a well known phenomenon that a lattice mismatch between substrate and film will result in a pseudomorphically strained layer with subsequent stress relieve by different accommodation mechanisms such as generation of misfit dislocations, stacking faults or undulations of the lattice planes or a combination of these. As a rule of thumb, a lattice mismatch in the order of 1-2 % is accommodated by the generation of stressed pseudomorphically grown films up to a critical thickness, $t_c$, which decreases with increasing lattice mismatch. For YBCO deposited onto $LaGaO_3$ single crystal substrates, $t_c$ has been determined to be around 50 nm [15]. A careful analysis of the pressure and strain dependence of $T_c$ for different HTS materials by Locquet et al. [9] shows that in most cases the uniaxial $dT_c/dp$ values have different signs and thus add up to an increase of $T_c$ for compressive strain and a decrease of $T_c$ for tensile strain. Indeed, Locquet et al. could demonstrate, that biaxial strain in $La_{1.9}Sr_{0.1}CuO_4$ (LSCO) thin films pushes $T_c$ up from the bulk value of 25K to 50 K.

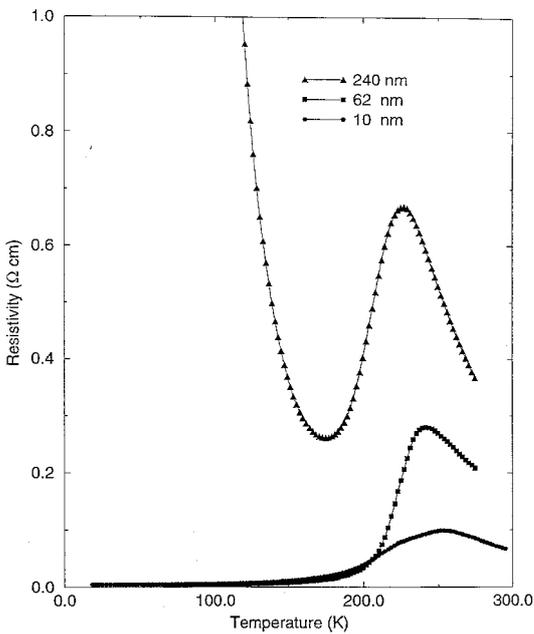

In the case of manganite thin films compressive strain can cause an even more dramatic effect on the temperature dependence of resis-tivity. In Fig. 3 the $\rho(T)$ curves $La_{.88}Sr_{.1}MnO_3$ thin films of different thickness are represented. Whereas the thick film (240 nm) shows the bulk-like behavior with a metal-insulator transition, $T_{MI}$, at 220K followed by a transition to a charged ordered insulator at 170K, charge ordering is destroyed in the homogeneously strained thin films. Compressive epitaxial strain causes here the transition from a ferromagnetic insulator to a ferromagnetic metal [4].

Fig. 3. Temperature dependence of resistivity of $La_{.88}Sr_{.1}MnO_3$ thin films of different thickness.

Similarly, in thin films of the charge-ordered [CO] $Pr_{1/2}Ca_{1/2}MnO_3$ the robustness of the CO state depends strongly on the strains and thus on the film thickness as shown by Prellier et al. [8].

## 5. YBCO thin films deposited on vicinal cut $SrTiO_3$ single crystals

When (001)-oriented $SrTiO_3$ substrates with an intentional miscut (angle $\alpha < 15^0$) towards the [010] direction are annealed at 950 $^0$C a regular step and terrace structure is generated with the step height of typically one unit cell of $SrTiO_3$ [a = 0.3905 nm] and a step width corresponding to w = a / tan$\alpha$. In contrast to films grown onto closely lattice matched (001)-oriented perovskite-type oxide substrates the films deposited onto the vicinal cut substrates show a terrace-like surface morphology with steps along the [100] direction, indicating a change from the usual Stranski–Krastanov growth to a step-flow growth mode. The substrate-mediated modification of the growth mode $YBa_2Cu_3O_{7-x}$ influences the micro-structure of the films and causes an artificially introduced anisotropy of their transport and pinning properties. We find a close correlation between the film morphology and the film properties as revealed by transport measurements and Raman spectroscopy. In Fig. 4 the surface of a $10^0$ miscut $SrTiO_3$ surface clearly demonstrates a remarkable nanoscale terrace structure as a result of the UHV annealing. Along the [010] direction regularly spaced terraces with a period of 2.3 nm and a step height of $\approx 0.39$ nm corresponding to one unit cell of $SrTiO_3$ is found. The terrace-like structure is basically preserved after YBCO film deposition, indicating a change of the usual Stranski–Krastanov island growth a step-flow growth for YBCO. Compared to the substrate, the terraces of the film are wider by roughly a factor of three ($\approx 7$ nm). The step heights, however, are multiples of 0.2 nm, deviating from the full integer of the unit cell height of 1.2 nm of YBCO. This implies that the unit cells grown on the upper and the lower part of a single unit cell step of the substrate can be shifted vertically forming an antiphase boundary (APB). The planar APBs are oriented perpendicular to the film plane forming a regular nanoscale array with an APB distance of around 7 nm. A detailed analysis of the defect structure due to the nanoscale surface step structure of the substrate and its implications to transport properties and flux pinning is given in by Haage et al. [16]. The predominant modifications of the properties of such films are summarized as follows [17]: (i) Strong anisotropy of the electrical dc resistivity; (ii) thickness dependent enhancement of the critical current; (iii) partial detwinning of the films with a better perfection of the CuO chains along the step edges; (iIV) anisotropy of the dimensionality of the fluctuation conductivity above $T_c$ with a large temperature range for the 3-dimensional fluctuations along the step edges and a small range for 3-dimensional fluctuations perpendicular to the step edges.

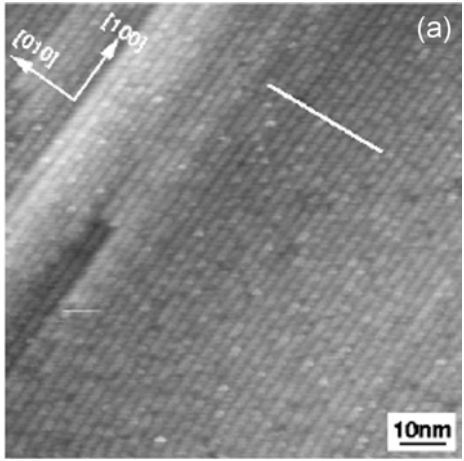

Fig. 4. STM image of a $10^0$ vicinal cut $SrTiO_3$ single crustal substrate after UHV annealing.

Similarly, manganite thin films grown on vicinal cut $SrTiO_3$ substrates show an in-plane magnetic anisotropy with the easy axes along the substrate steps. Over a large angular range the angular dependence of the magnetic switching field is found to obey the $1/\cos\theta$ law, indicating that the magnetic reversal is completed by a $180^0$ domain nucleation and sweeping along the easy axis [18].

## 6   Conclusions and outlook

The experiments described above demonstrate the new possibilities in ceramic thin film research if the substrate is not only treated as a support material for the films but also additionally regarded as a functional integrated part of the system film/substrate. Making use of epitaxial strain opens the possibility to externally affect the atom arrangement and thus the properties of the films. Modifying the growth mode from Stranski–Krastanov to step flow in the case of the vicinal cut substrates is a new possibility for tailoring the defect structure and thus flux-line pinning sites in cuprates and in-plane magnetic anisotropy in manganites. Recently, a further technique for controlled substrate surface modification has been introduced using either ion implantation with a focused ion beam microscope [19] or laser surface treatment to regularly etch µm scale grooves or trenches into the substrate. The physical concept behind these experiments is the search of matching effects in HTS flux-line pinning, fabrication of flux guides and formation of regular arrays of manganite ferromagnetic quantum dots for in plane spin valve devices.

**References**

[1]  Interfaces in High $T_c$ Superconducting Systems, Springer Berlin, (1993).  S.L. Shinde; A. Rudman (eds.),
[2]  J. M. Philips in H. Weinstock and R. W. Ralston edts. The New Superconducting Electronics NATO ASI Series, Kluwer, Doordrecht/Boston/London, p. 59 (1993)
[3]  T. H. Hylton and M. R. Beasley, Phys. Rev. B **41** 11669 (1990).
[4]  F.S. Razavi, M. Gross, H.-U. Habermeier, O.Lebedev, S. Amelinckx, G. V. van Tendeloo, A. Vigliante, Appl. Phys. Lett. **76,** 155 (2000).
[5]  H.-U. Habermeier, F.S. Razavi, O. Lebedev, G.M. Gross, R. Praus, and P.X. Zhang phys. stat.sol.b **215** , 679 (1999).
[6]  A. Ramirez, J. Phys. Condens. Matter **9** , 8171 (1997).
[7]  J. Coey,  M. Viret, and S. Von Molnar, Adv. Phys. **48**, 167 (1999).
[8]  W. Prellier, Ch. Simon, A. M. Haghiri-Gosnet, B. Mercey, and B. Raveau, Phys. Rev. B **62**, R16337 (2000)
[9]  J.-P. Locquet, J.  Perret, J. Fompeyrine, E. Machler,J. W. Seo, and G. V. van Tendeloo, Nature **394** , 453 ( 1998).
[10] E.Pavarini, I. Dasgupta, T. Sasha-Dasgupta, OI. Jepsen, and O. K. Andersen, Phys. Rev. Lett. 87, 47003 (2001).
[11]  R. Feenstra, T. B. Lindemer, J. D. Budai, M. D. Galloway, J. Appl. Phys. **69** , 6569  (1991).
[12] I. D. Raistrick, M. Hawley, M., in: S.L. Shinde; A. Rudman (eds.),   Interfaces in High $T_c$ Superconducting Systems, Springer Berlin, p. 28  (1993).
[13] R. E. Somekh, Z. H. Barber, J. Evetts, . in: J. Evetts, J. (ed.)  Concise Encyclopedia of Magnetic and Superconducting Materials, Pergamon Press Oxford, p. 431 (1992).
[14] O. I. Lebedev, G. van Tendeloo, S. Amelinckx, B. Leibold and H.-U. Habermeier, Phys. Rev. B. **58**, 8065 (1998).
[15] M. Ece, E. Garcia-Gonzalez, H.-U.Habermeier, and B. Oral, J. Appl. Phys .**77,** 1646 (1995).
[16] T. Haage, J. Zegenhagen, J. Q. Li, H.-U. Habermeier, M. Cardona, Ch. Jooss, R. Warthmann, A. Forkl, and H. Kronmüller, Phys. Rev. B **56**, 8404 (1997).
[17] H.-U. Habermeier,  Proc. SPIE **3481**, 204 (1998).
[18] Z.-H. Wang, G. Cristiani and H.-U. Habermeier, Appl. Phys. Lett. **82**, 3731 (2003)
[19] J. Albrecht, S. Leonhardt, R. Spolanek, U. Täffner, H.-U. Habermeier, and G. Schütz,  Surface Science **547**, L847 (2003).